\documentclass[oldversion]{aa}    

\usepackage{graphicx}
\usepackage{txfonts}
\usepackage{natbib}

\bibpunct{(}{)}{;}{a}{}{,}

\begin{document}

\title{Sub-arcsecond CO(1--0) and CO(2--1) observations of the ultraluminous infrared galaxy IRAS 10190+1322\thanks{Based on observations carried out with the IRAM Plateau de Bure Interferometer. IRAM is supported by INSU/CNRS (France), MPG (Germany) and IGN (Spain).}}

\author{J. Graci{\'a}-Carpio\inst{1} \and P. Planesas\inst{1} \and L. Colina\inst{2}}

\offprints{J. Graci{\'a}-Carpio, \email{j.gracia@oan.es}}

\institute{Observatorio Astron{\'o}mico Nacional (OAN), Observatorio de Madrid, Alfonso XII 3, 28014 Madrid, Spain\\
           \email{j.gracia@oan.es, p.planesas@oan.es}
           \and
           Consejo Superior de Investigaciones Cient{\'{\i}}ficas (CSIC), Instituto de Estructura de la Materia, Serrano 121, 28006 Madrid, Spain\\
           \email{colina@damir.iem.csic.es}}

\date{Received 9 March 2007; accepted 10 April 2007}

\titlerunning{Sub-arcsecond CO observations of IRAS 10190+1322}

\authorrunning{J. Graci{\'a}-Carpio et al.}

\abstract{We present the results of high resolution mapping of the CO(1--0) and CO(2--1) emission of the ultraluminous infrared galaxy (ULIRG) IRAS 10190+1322, with the IRAM interferometer, down to an angular resolution of $\sim$0.3$\arcsec$. This object is composed of two interacting galaxies with a projected nuclear separation of 6\,kpc, and was selected to analyze the physical and dynamical properties of the molecular gas in each galaxy in order to study the conditions that lead a galaxy pair to become ultraluminous in the infrared. With the exception of Arp 220, the closest ULIRG, this is the first time that the CO emission is morphologically and kinematically resolved in the two interacting galaxies of a ULIRG system. In one of the galaxies the molecular gas is highly concentrated, distributed in a circumnuclear disk of 1.7\,kpc in size. The molecular gas in the presumably less infrared luminous galaxy is distributed in a more extended disk of 7.4\,kpc. The molecular gas mass accounts for $\sim$10\% of the dynamical mass in each galaxy. Both objects are rich enough in molecular gas, $M_{\rm gas} \sim 4\;10^{9}$\,M$_{\sun}$, as to experience an infrared ultraluminous phase.}


\maketitle

\section{Introduction\label{Introduction}}

Ultraluminous infrared galaxies (ULIRGs: $L_{\rm ir}$\footnote{For the definition of $L_{\rm ir}$ see Table 1 in \citet{Sanders96}. In this paper we will assume a flat $\rm{\Lambda}$-dominated cosmology described by $\rm{H_{0}} = 71\,km\,s^{-1}\,\rm{Mpc}^{-1}$ and $\rm{\Omega_{m}} = 0.27$ \citep{Spergel03}.} $\geq 10^{12}$\,L$_{\sun}$) present signs of current or past interactions in their optical and near-infrared images in a fraction that approaches 100\% \citep{Murphy96,Bushouse02,Veilleux02}. The ULIRG phenomenon has been observed along the full interaction phase, from widely separated pairs of galaxies (up to a projected nuclear separation $S_{\rm N} \sim 50$\,kpc) to advanced and relaxed mergers. However, the distribution of nuclear separations is highly peaked at low values with 60\% of the IRAS 1\,Jy ULIRG sample having $S_{\rm N} < 2.5$\,kpc \citep{Veilleux02}. In particular, this has hindered the study of the molecular gas properties in ULIRGs, sub-arcsecond resolution observations being necessary to morphologically and kinematically resolve the emission of the two interacting galaxies. In fact, this has been achieved only in \object{Arp 220}, the most nearby ULIRG \citetext{\citealt{Downes98}, hereafter\defcitealias{Downes98}{DS98}\citetalias{Downes98}; \citealt{Sakamoto99}}.

To date, there have been two millimeter interferometric surveys devoted to study the molecular gas content of ULIRGs with separated pairs of galaxies, but they have lacked the necessary sensitivity and/or angular resolution to detect and resolve the CO emission of the two galaxies. \citet{Dinh-V-Trung01} detected CO(1--0) emission with the BIMA array in five out of six targets (20\,kpc $< S_{\rm N} <$  51\,kpc), but only in one component of the ULIRG pairs. Interferometric observations with the OVRO array of five double nuclei ULIRGs (3\,kpc $< S_{\rm N} <$ 5\,kpc) by \citet{Evans02} have succeeded in detecting the CO(1--0) emission of both nuclei in two cases, but did not resolve their individual molecular gas distributions. The overall conclusion of all these observations, mostly derived from upper limits, was that $>$2/3 of the molecular gas mass lies around one of the nuclei, being this nucleus the most active of the two (LINER or Seyfert optical classification). Additional observations are required to confirm or reject this conclusion. The recent upgrade of the IRAM Plateau de Bure Interferometer (PdBI) to larger baselines currently allows this sort of studies.

\begin{figure*}[ht!]
\centering
\resizebox{0.89\hsize}{!}{\includegraphics{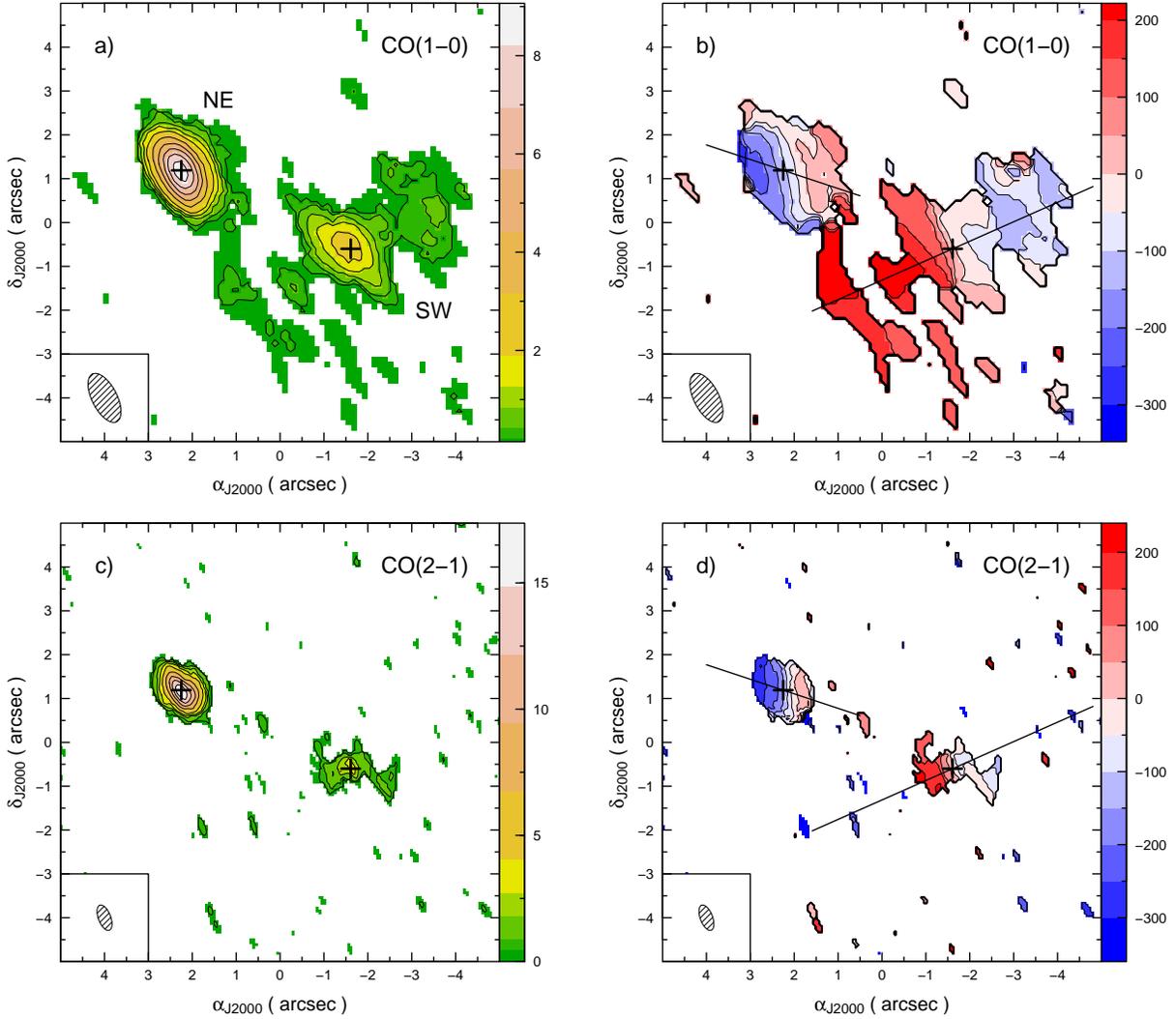}} 
\caption{{\bf a)} CO(1--0) integrated intensity map of IRAS 10190+1322 at the reference position ($\alpha_{\rm J2000}$, $\delta_{\rm J2000}$) = ($10^{\rm h}21^{\rm m}42\fs6$, $13\degr06\arcmin54\farcs4$). Contours for the CO integrated intensity are $1 \sigma$, $2 \sigma$, $4 \sigma$, $6 \sigma$ and $9 \sigma$ to $39 \sigma$ in steps of $6 \sigma$ ($1 \sigma = 0.21$ Jy km\,s$^{-1}$ beam$^{-1}$). The synthesized beam size, shown in the bottom left corner, is $1.22\arcsec \times 0.57\arcsec$ with a position angle of $27\degr$. The black crosses mark the central position of the NE and SW galaxies (see Table~\ref{Table}) determined by elliptical Gaussian fits in the uv plane. {\bf b)} CO(1--0) velocity map relative to the redshift of observation ($z_{\rm obs} = 0.07656$). Contours for the CO velocity are $-300$ to 200\,km\,s$^{-1}$ in steps of 50\,km\,s$^{-1}$. The straight lines indicate the position angle of the mayor axis in each galaxy estimated from the velocity channel maps. {\bf c)} CO(2--1) integrated intensity map. Contours for the CO integrated intensity are $1 \sigma$, $2 \sigma$, $4 \sigma$, $6 \sigma$ and $9 \sigma$ to $33 \sigma$ in steps of $6 \sigma$ ($1 \sigma = 0.45$ Jy km\,s$^{-1}$ beam$^{-1}$). The synthesized beam size is $0.61\arcsec \times 0.28\arcsec$ with a position angle of $21\degr$. {\bf d)} CO(2--1) velocity map. Contours for the CO velocity are $-300$ to 200\,km\,s$^{-1}$ in steps of 50\,km\,s$^{-1}$. In order to show the more extended and weaker emission coming from the SW galaxy, the moment maps have been derived applying a 3$\sigma$--clipping to each channel of the CO(1--0) and CO(2--1) clean maps in the velocity range between $-360$ and 240\,km\,s$^{-1}$.} 
\label{Maps}
\end{figure*}

\object{IRAS 10190+1322} ($L_{\rm ir} = 10^{12}$\,L$_{\sun}$ at $D_{\rm L} = 340$\,Mpc) is composed of two interacting galaxies with a projected nuclear separation of 6\,kpc ($4\arcsec$). The south-west (SW) galaxy is brighter than the north-east (NE) galaxy in the optical but the latter is brighter in the near-infrared \citep{Murphy01,Kim02,Imanishi06,Dasyra06}. At 15\,GHz \citet{Nagar03} detected compact radio continuum emission only in the NE galaxy. The system was not spatially resolved at 4.86\,GHz and 1.4\,GHz by \citet{Crawford96} and \citet{White97}, but the radio continuum distributions peak towards the NE galaxy position. All these observations point to the NE galaxy to host the bulk of the mid- and far-infrared luminosity of IRAS 10190+1322. However, \citet{Imanishi07} found that the Spitzer 5.2-14.5\,$\mu$m spectra of the NE galaxy is only slightly brighter than that of the SW galaxy. \citet{Murphy01} obtained Pa$\alpha$ images of both galaxies and concluded that the emission in the NE galaxy is more compact and absorbed than in the SW galaxy. This result agrees with their kinematical analysis where the two galaxies are approaching to experience a second encounter after the first close passage, and the NE galaxy has resulted more affected by the interaction than the SW galaxy due to the initial configuration of the encounter. \citet{Rupke02} classified the NE galaxy as a LINER and the SW galaxy as an HII galaxy. \citet{Farrah03} modelled the overall spectral energy distribution (SED) of the system and determined that the contribution to $L_{\rm ir}$ from an active galactic nucleus (AGN) could be as high as the 20\%. \citet{Imanishi06,Imanishi07} did not find evidences of an obscured AGN in the $L$-band and 5.2-14.5\,$\mu$m spectra of both galaxies.

\section{Observations\label{Observations}}

IRAS 10190+1322 ($z = 0.07656$) was observed with the six antennae of the PdBI in February 2006 in the new extended A configuration (baselines up to 760\,m). The 3\,mm and 1\,mm receivers were tuned to the CO(1--0) and CO(2--1) redshifted lines at 107.074\,GHz and 214.143\,GHz, respectively. The system temperatures were 140\,K for CO(1--0) and 250\,K for CO(2--1). Four correlator units covered a total bandwidth of 580\,MHz at each frequency. Phase and amplitude calibrations were done on nearby quasars. 

Reduction using the GILDAS software provided data cubes with a resolution of 0.12$\arcsec/$pixel and 0.06$\arcsec/$pixel that were cleaned with the standard method with a velocity resolution of 30\,km\,s$^{-1}$. 
The restored clean beams are $1.22\arcsec \times 0.57\arcsec$ (PA $= 27\degr$) at 2.8\,mm and $0.61\arcsec \times 0.28\arcsec$ (PA $= 21\degr$) at 1.4\,mm. The rms noise levels in the cleaned maps (at 30\,km\,s$^{-1}$ velocity resolution) are 1.57\,mJy\,beam$^{-1}$ for the CO(1--0) line and 4.25\,mJy\,beam$^{-1}$ for the CO(2--1) line. We tentatively detected 2.8\,mm continuum emission of 0.9\,mJy, just below the 3$\sigma$ level, towards the central position of the NE galaxy. This density flux is compatible with the 60\,$\mu$m and 100\,$\mu$m IRAS density fluxes of the whole system assuming optically thin grey-body emission with a dust temperature $T_{\rm dust} \sim 38$\,K and a spectral index of the dust absorption coefficient $\beta \sim 1.4$. 

IRAS 10190+1322 was also observed with the IRAM 30 meter telescope in June 2006. We used 2 SIS 3\,mm receivers to observe the 2 polarizations simultaneously at 107.074\,GHz. At this frequency, the telescope half-power beam width was 23$\arcsec$ and the antenna temperature to flux conversion factor was $S/T_{\rm a}^*$ = 6.3\,Jy$/$K. During the observations the typical system temperature was 170\,K (on the $T_{\rm a}^*$ scale). The observations were done in wobbler switching mode, with reference positions offset by 2$\arcmin$ in azimuth. Two 1\,MHz filter banks provided a total bandwidth of 512\,MHz, or about 1400\,km\,s$^{-1}$.

\begin{figure}
\centering
\resizebox{\hsize}{!}{\includegraphics{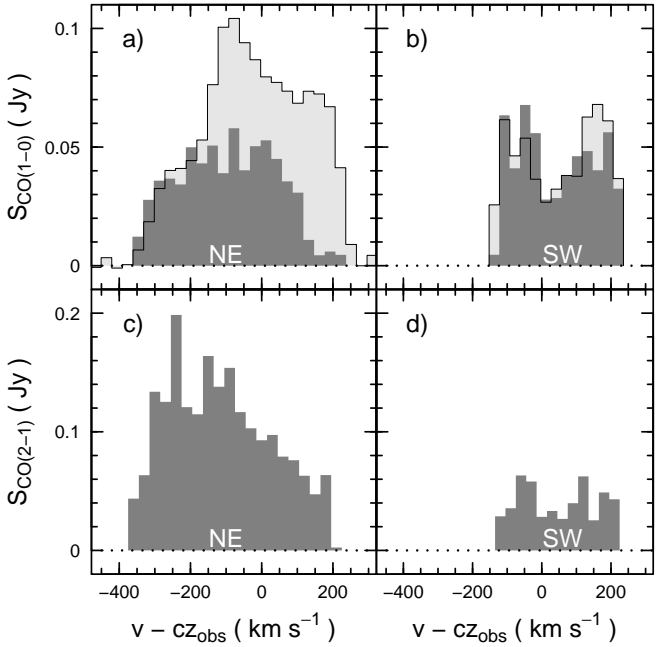}} 
\caption{{\bf a)} IRAM 30 meter telescope CO(1--0) spectrum of IRAS 10190+1322 complex (light grey). The CO(1--0) line profile of the NE galaxy observed with the PdBI is represented in dark grey at the same panel. {\bf b)} The result of the substraction of the NE galaxy PdBI CO(1--0) line to the total single dish CO(1--0) line is shown in light grey. In dark grey we plot the PdBI CO(1--0) line spectrum of the SW galaxy. {\bf c)} and {\bf d)} PdBI CO(2--1) line profiles of the NE and SW galaxies, respectively. The PdBI spectra of each galaxy were derived integrating inside the area that includes all pixels with an emission $>$3$\sigma$ in any of the 30\,km\,s$^{-1}$ channel maps.}
\label{Lines}
\end{figure}

\section{Results\label{Results}}

\begin{figure*}
\centering
\resizebox{\hsize}{!}{\includegraphics{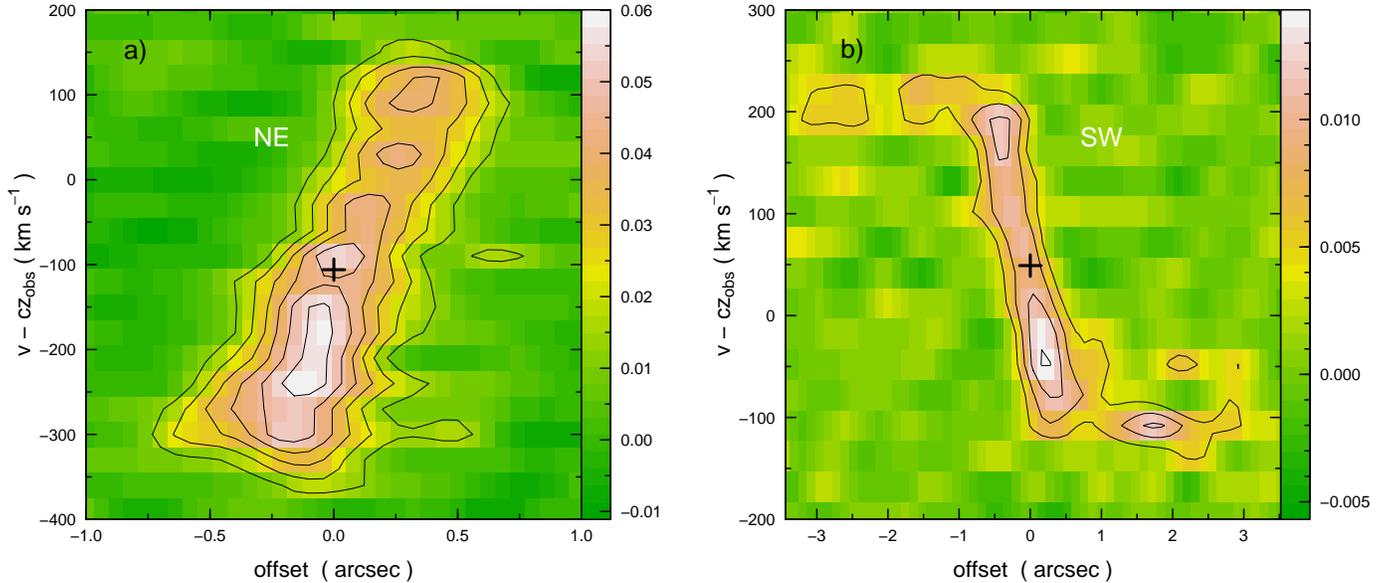}} 
\caption{{\bf a)} CO(2--1) position-velocity diagram of the NE galaxy along the direction of its major axis (see Fig.~\ref{Maps}d) at a velocity resolution of 30\,km\,s$^{-1}$. Contours for the CO intensity are $3\sigma$ to $13\sigma$ in steps of $2\sigma$ ($1\sigma = 4.25$\,mJy\,beam$^{-1}$). The black cross marks the central position and velocity of the NE galaxy as measured in the CO(1--0) map. {\bf b)} CO(1--0) position-velocity diagram of the SW galaxy along the direction of its major axis (see Fig.~\ref{Maps}b) at a velocity resolution of 30\,km\,s$^{-1}$. Contours for the CO intensity are $3\sigma$ to $9\sigma$ in steps of $2\sigma$ ($1\sigma = 1.57$\,mJy\,beam$^{-1}$). The black cross marks the central position and velocity of the SW galaxy as measured in the CO(1--0) map.}
\label{Slices}
\end{figure*}

Emission of the first two CO rotational transitions has been detected and spatially resolved towards both galaxies. A summary of the observational results is given in Table~\ref{Table}. In Fig.~\ref{Maps} we have plotted the integrated intensity and velocity maps of the two lines. CO(1--0) and CO(2--1) spectra are shown in Fig.~\ref{Lines} and the position-velocity diagrams along the major axis of each galaxy are in Fig.~\ref{Slices}.

\begin{table}
\caption{IRAS 10190+1322 main properties}
\label{Table}
\centering
\begin{tabular}{@{}ll@{}cc@{}}
\hline
\hline
\\[-0.2cm]
                                    & Unit            &             NE Galaxy             &             SW Galaxy             \\[0.1cm]
\hline
\\[-0.2cm]
$\alpha_{\rm J2000}$                & \dots           & $10^{\rm h}21^{\rm m}42\fs754$(1) & $10^{\rm h}21^{\rm m}42\fs490$(4) \\[0.05cm]
$\delta_{\rm J2000}$                & \dots           &  $13\degr06\arcmin55\farcs59$(1)  &  $13\degr06\arcmin53\farcs80$(5)  \\[0.05cm]
$V - {\rm c}\,z_{\rm obs}$          & km\,s$^{-1}$    &            $-100$(10)             &                40(10)             \\[0.05cm]
$I_{\rm CO(1-0)}$                   & Jy km\,s$^{-1}$ &                20(1)              &                17(1)              \\[0.05cm]
$I_{\rm CO(2-1)}$                   & Jy km\,s$^{-1}$ &                62(3)              &                15(2)              \\[0.05cm]
$\Delta V_{\rm CO(1-0)}$\,(FWHM)    & km\,s$^{-1}$    &               450(30)             &               360(30)             \\[0.05cm]
$L'_{\rm CO(2-1)}/L'_{\rm CO(1-0)}$ & \dots           &              0.78(5)              &               0.6(1)$^{\rm a}$    \\[0.05cm]
$R_{\rm CO}$                        & kpc             &               0.9(1)              &               3.7(3)              \\[0.05cm]   
$D_{\rm major}/D_{\rm minor}$       & \dots           &               1.3(1)              &               1.4(2)              \\[0.05cm]   
$M_{\rm gas}$                       & M$_{\sun}$      &     4.2(2)$\,\times\,10^{9}$      &     3.6(2)$\,\times\,10^{9}$      \\[0.05cm]
$M_{\rm dyn}$                       & M$_{\sun}$      &       4(1)$\,\times\,10^{10}$     &       5(2)$\,\times\,10^{10}$     \\[0.05cm]
$M_{\rm gas}/M_{\rm dyn}$           & \dots           &              0.11(3)              &              0.07(3)              \\[0.1cm]
\hline 
\\[-0.2cm]
\end{tabular}
\begin{minipage}{0.48 \textwidth}
\footnotesize{CO luminosities have been computed using equation~3 in \citet{Solomon97}. Numbers in parentheses indicate estimated measurement errors in units of the last significant figures. $^{\rm a}$ Corresponding to the central region ($R_{\rm CO} = 1.1$\,kpc) of the SW galaxy.}
\end{minipage}
\end{table}

The CO distribution of the NE galaxy is highly concentrated (Figs.~\ref{Maps}a and \ref{Maps}c).
Given its compactness, it is not surprising that $\sim$100\% of the CO(1--0) emission of this galaxy is recovered by the interferometer (Fig.~\ref{Lines}a). Very likely most of the CO(2--1) emission is also recovered, as the $L'_{\rm CO(2-1)}/L'_{\rm CO(1-0)}$ luminosity ratio is $\sim$0.78, similar to the typical luminosity ratios found in normal spiral galaxies \citep{Braine93}. 

The emission is resolved in both transitions, but it is in the CO(2--1) intensity map, due to the smaller synthesized beam, that the structure of the galaxy is more clearly seen. The centroid of the molecular gas distribution of the NE galaxy coincides with its 15\,GHz radio continuum core \citep{Nagar03} within 0.1$\arcsec$. The deconvolved major axis of the CO emitting region (at zero intensity) is 1.7\,kpc (1.2$\arcsec$), similar to the sizes found in ULIRGs in advanced mergers \citepalias{Downes98}. From the ratio between the major and minor axis of the CO distribution we derive an inclination for the galaxy of $i \sim$40$\degr$. 

The symmetric CO(1--0) line profile and the resolved distribution of the CO(2--1) intensity map (Figs.~\ref{Lines}a and ~\ref{Maps}c) hint that the molecular gas is distributed in a circumnuclear disk or ring. If such were the case, the asymmetric CO(2--1) line profile (Fig.~\ref{Lines}c) could be explained by a differential excitation or by some missing flux at the velocity range between 0 to 200\,km\,s$^{-1}$. 
The CO(2--1) position-velocity diagram along the mayor axis of the galaxy shows a linear increment of the radial velocity as a function of the distance to the center up to $R_{\rm CO}=0.6\arcsec$ (Fig.~\ref{Slices}a). The terminal rotational velocity at this distance from the nucleus is $\sim$275\,km\,s$^{-1}$ (cf. Figs. \ref{Lines}a and \ref{Slices}a). Therefore, we derive a dynamical mass corrected from inclination of $4\;10^{10}$\,M$_{\sun}$. The molecular gas mass in the same region estimated from the CO(1--0) flux, and assuming a conversion factor $\alpha = 0.8$\,M$_{\sun}$\,(K\,km\,s$^{-1}$\,pc$^{2}$)$^{-1}$ suitable for ULIRGs \citepalias{Downes98}, is $M_{\rm gas} = 4.2\;10^9$\,M$_{\sun}$, i.e. 11\%\ of the dynamical mass. 

The molecular gas distribution of the SW galaxy is more extended than that of the NE galaxy, with a compact peak that remains detected even at the highest resolution. The deconvolved size of the major axis of the large CO(1--0) emitting region is 7.4\,kpc. From the ratio between the major and minor axis of the CO distribution we derive an inclination for the galaxy of $i \sim 45\degr$. In Fig.~\ref{Lines}b we compare the deduced single beam CO(1--0) spectrum of the SW galaxy with the CO(1--0) line emission detected by the PdBI. We estimate that $\sim$100\% of the CO(1--0) emission of the galaxy is recovered by the interferometer. However, probably most of the extended CO(2--1) line emission of the galaxy is lost due to the higher frequency of the CO(2--1) transition. In fact the $L'_{\rm CO(2-1)}/L'_{\rm CO(1-0)}$ luminosity ratio calculated in the central 2.2\,kpc of the galaxy is $\sim$0.6, a low value when compared to those found in spiral galaxies.

The double horned CO(1--0) and CO(2--1) line profiles of the SW galaxy (Figs.~\ref{Lines}b and \ref{Lines}d) hint that the molecular gas is distributed in a circumnuclear disk or ring. The position-velocity diagram along the mayor axis of the galaxy in the CO(1--0) transition is plotted in Fig.~\ref{Slices}b. The radial velocity increases as a function of the distance to the center up to $R_{\rm CO}=0.8\arcsec$ and then remains constant up to $R_{\rm CO}=2.6\arcsec$ (3.7\,kpc). The terminal rotational velocity is $\sim$170\,km\,s$^{-1}$ (cf. Figs.~\ref{Lines}b and \ref{Slices}b), therefore we derive a dynamical mass of $5\;10^{10}$\,M$_{\sun}$ within this radius. The molecular gas mass in the same region is $M_{\rm gas} = 3.6\;10^9$\,M$_{\sun}$ and $M_{\rm gas}/M_{\rm dyn} \sim7\%$.

\section{Discussion and conclusions\label{Discussion and conclusions}}

From our CO(1--0) interferometric observations of IRAS 10190+1322 we conclude that the two galaxies of the system have similar CO(1--0) luminosities, i.e. similar gas reservoirs. The main difference between them is the more compact molecular gas distribution of the NE galaxy, which is a consequence of the initial configuration of the encounter \citetext{see \citealt{Murphy01} for a detailed discussion}, being this galaxy the most affected by the interaction. The NE galaxy is classified as a LINER and the SW galaxy as a HII galaxy by \citet{Rupke02}. Therefore, contrary to the \cite{Dinh-V-Trung01} and \cite{Evans02} conclusions, we find that the active galaxy does not have the larger molecular gas amount, but a higher concentration of gas in its nuclear region.

\citet{Farrah03} obtained from starburst+AGN model fitting of IRAS 10190+1322 SED a total star formation rate of 195\,M$_{\sun}$\,yr$^{-1}$. At this rate the total molecular gas reservoir of the NE galaxy will be exhausted in $\sim$20\,Myr, considering that the bulk of the star formation takes place in this galaxy (see Sect.~\ref{Introduction} and the discussion of our 2.8\,mm continuum detection in Sect.~\ref{Observations}). This period is shorter than the time needed for the two galaxies to coalesce or experience a second close passage ($\gtrsim$ 50\,Myr), assuming a transversal velocity of the order of the measured relative radial velocity ($\sim$140\,km\,s$^{-1}$, see Table~\ref{Table}). The second encounter or final merger of the system may result in the concentration of the molecular gas in the central region of the SW galaxy and the restarting of the ULIRG phase with a new starbursting event.

\begin{acknowledgements}
We thank M. Imanishi for fruitful discussion about the relative mid- and far-infrared luminosity of IRAS 10190+1322 nuclei and S. Garc{\'{\i}}a-Burillo, R. Neri and F. Colomer for their help with the calibration and interpretation of our CO maps. We would also like to thank the referee, A. Beelen, for his useful comments which helped to improve the final version of the article. This work has been partially supported by the Spanish MEC and Feder funds under grant ESP2003-04957 and by SEPCT/MEC under grants AYA2003-07584 and AYA2002-01055.
\end{acknowledgements}

\bibliographystyle{aa} 

\bibliography{main.bib}

\begin{thebibliography}{21}
\expandafter\ifx\csname natexlab\endcsname\relax\def\natexlab#1{#1}\fi

\bibitem[{{Braine} {et~al.}(1993){Braine}, {Combes}, {Casoli}, {Dupraz},
  {Gerin}, {Klein}, {Wielebinski}, \& {Brouillet}}]{Braine93}
{Braine}, J., {Combes}, F., {Casoli}, F., {et~al.} 1993, \aaps, 97, 887

\bibitem[{{Bushouse} {et~al.}(2002){Bushouse}, {Borne}, {Colina}, {Lucas},
  {Rowan-Robinson}, {Baker}, {Clements}, {Lawrence}, \& {Oliver}}]{Bushouse02}
{Bushouse}, H.~A., {Borne}, K.~D., {Colina}, L., {et~al.} 2002, \apjs, 138, 1

\bibitem[{{Crawford} {et~al.}(1996){Crawford}, {Marr}, {Partridge}, \&
  {Strauss}}]{Crawford96}
{Crawford}, T., {Marr}, J., {Partridge}, B., \& {Strauss}, M.~A. 1996, \apj,
  460, 225

\bibitem[{{Dasyra} {et~al.}(2006){Dasyra}, {Tacconi}, {Davies}, {Genzel},
  {Lutz}, {Naab}, {Burkert}, {Veilleux}, \& {Sanders}}]{Dasyra06}
{Dasyra}, K.~M., {Tacconi}, L.~J., {Davies}, R.~I., {et~al.} 2006, \apj, 638,
  745

\bibitem[{{Dinh-V-Trung} {et~al.}(2001){Dinh-V-Trung}, {Lo}, {Kim}, {Gao}, \&
  {Gruendl}}]{Dinh-V-Trung01}
{Dinh-V-Trung}, {Lo}, K.~Y., {Kim}, D.-C., {Gao}, Y., \& {Gruendl}, R.~A. 2001,
  \apj, 556, 141

\bibitem[{{Downes} \& {Solomon}(1998)}]{Downes98}
{Downes}, D. \& {Solomon}, P.~M. 1998, \apj, 507, 615

\bibitem[{{Evans} {et~al.}(2002){Evans}, {Mazzarella}, {Surace}, \&
  {Sanders}}]{Evans02}
{Evans}, A.~S., {Mazzarella}, J.~M., {Surace}, J.~A., \& {Sanders}, D.~B. 2002,
  \apj, 580, 749

\bibitem[{{Farrah} {et~al.}(2003){Farrah}, {Afonso}, {Efstathiou},
  {Rowan-Robinson}, {Fox}, \& {Clements}}]{Farrah03}
{Farrah}, D., {Afonso}, J., {Efstathiou}, A., {et~al.} 2003, \mnras, 343, 585

\bibitem[{{Imanishi} {et~al.}(2007){Imanishi}, {Dudley}, {Maiolino}, {Maloney},
  {Nakagawa}, \& {Risaliti}}]{Imanishi07}
{Imanishi}, M., {Dudley}, C.~C., {Maiolino}, R., {et~al.} 2007,
  astro-ph/0702136

\bibitem[{{Imanishi} {et~al.}(2006){Imanishi}, {Dudley}, \&
  {Maloney}}]{Imanishi06}
{Imanishi}, M., {Dudley}, C.~C., \& {Maloney}, P.~R. 2006, \apj, 637, 114

\bibitem[{{Kim} {et~al.}(2002){Kim}, {Veilleux}, \& {Sanders}}]{Kim02}
{Kim}, D.-C., {Veilleux}, S., \& {Sanders}, D.~B. 2002, \apjs, 143, 277

\bibitem[{{Murphy} {et~al.}(1996){Murphy}, {Armus}, {Matthews}, {Soifer},
  {Mazzarella}, {Shupe}, {Strauss}, \& {Neugebauer}}]{Murphy96}
{Murphy}, Jr., T.~W., {Armus}, L., {Matthews}, K., {et~al.} 1996, \aj, 111,
  1025

\bibitem[{{Murphy} {et~al.}(2001){Murphy}, {Soifer}, {Matthews}, \&
  {Armus}}]{Murphy01}
{Murphy}, Jr., T.~W., {Soifer}, B.~T., {Matthews}, K., \& {Armus}, L. 2001,
  \apj, 559, 201

\bibitem[{{Nagar} {et~al.}(2003){Nagar}, {Wilson}, {Falcke}, {Veilleux}, \&
  {Maiolino}}]{Nagar03}
{Nagar}, N.~M., {Wilson}, A.~S., {Falcke}, H., {Veilleux}, S., \& {Maiolino},
  R. 2003, \aap, 409, 115

\bibitem[{{Rupke} {et~al.}(2002){Rupke}, {Veilleux}, \& {Sanders}}]{Rupke02}
{Rupke}, D.~S., {Veilleux}, S., \& {Sanders}, D.~B. 2002, \apj, 570, 588

\bibitem[{{Sakamoto} {et~al.}(1999){Sakamoto}, {Scoville}, {Yun}, {Crosas},
  {Genzel}, \& {Tacconi}}]{Sakamoto99}
{Sakamoto}, K., {Scoville}, N.~Z., {Yun}, M.~S., {et~al.} 1999, \apj, 514, 68

\bibitem[{{Sanders} \& {Mirabel}(1996)}]{Sanders96}
{Sanders}, D.~B. \& {Mirabel}, I.~F. 1996, \araa, 34, 749

\bibitem[{{Solomon} {et~al.}(1997){Solomon}, {Downes}, {Radford}, \&
  {Barrett}}]{Solomon97}
{Solomon}, P.~M., {Downes}, D., {Radford}, S.~J.~E., \& {Barrett}, J.~W. 1997,
  \apj, 478, 144

\bibitem[{{Spergel} {et~al.}(2003){Spergel}, {Verde}, {Peiris}, {Komatsu},
  {Nolta}, {Bennett}, {Halpern}, {Hinshaw}, {Jarosik}, {Kogut}, {Limon},
  {Meyer}, {Page}, {Tucker}, {Weiland}, {Wollack}, \& {Wright}}]{Spergel03}
{Spergel}, D.~N., {Verde}, L., {Peiris}, H.~V., {et~al.} 2003, \apjs, 148, 175

\bibitem[{{Veilleux} {et~al.}(2002){Veilleux}, {Kim}, \&
  {Sanders}}]{Veilleux02}
{Veilleux}, S., {Kim}, D.-C., \& {Sanders}, D.~B. 2002, \apjs, 143, 315

\bibitem[{{White} {et~al.}(1997){White}, {Becker}, {Helfand}, \&
  {Gregg}}]{White97}
{White}, R.~L., {Becker}, R.~H., {Helfand}, D.~J., \& {Gregg}, M.~D. 1997,
  \apj, 475, 479

\end{thebibliography}

\end{document}